\documentclass[aps,prl,twocolumn,showpacs,groupedaddress]{revtex4}
\usepackage{graphicx}

\begin{document}

\title{ Theory of plasmon-enhanced high-harmonic generation in the vicinity of
 metal nanostructures in noble gases}
\author{A. Husakou}
\email{gusakov@mbi-berlin.de}
\affiliation{Max Born Institute of Nonlinear Optics and Short Pulse Spectroscopy, Max
Born Str 2a, D-12489 Berlin, Germany}
\author{S.-J. Im}
\affiliation{Max Born Institute of Nonlinear Optics and Short Pulse Spectroscopy, Max
Born Str 2a, D-12489 Berlin, Germany}
\author{J. Herrmann}
\affiliation{Max Born Institute of Nonlinear Optics and Short Pulse Spectroscopy, Max
Born Str 2a, D-12489 Berlin, Germany}
\date{\today}

\begin{abstract}
 We present a semiclassical model for
plasmon-enhanced high-harmonic generation (HHG) 
in the vicinity of metal nanostructures.
 We show that both the inhomogeneity of the 
enhanced local fields and
 electron absorption by the metal surface 
play an important role in the HHG process and lead to
the generation of even harmonics and to a
significantly increased cutoff.
For the examples of silver-coated nanocones and bowtie antennas 
 we predict that the required intensity reduces by up
to three orders of magnitudes and the HHG cutoff increases by
more than a factor of two. 
The study of the enhanced high-harmonic generation is connected with
a finite-element simulation of the electric field enhancement due to the excitation
of the plasmonic modes.
\end{abstract}

\pacs{42.65.Ky, 78.67.Bf}
\maketitle

Progress of contemporary ultrafast laser physics has led to the
generation of attosecond laser pulses by high-harmonic generation
(HHG) \cite{hent} opening the way to exciting opportunities to study
time-resolved fundamental electronic processes in atoms and
molecules, and to investigate the collective and correlated dynamics
of many-electron systems on the attosecond scale
\cite{corkum,paul,zhang,kapteyn,kling,krausz,sansone}.
On the other
hand, in an another rapidly evolving field metallic
nanostructures can be used to achieve localization of light on
sub-wavelength nanometer scale and to realize huge plasmonic field
enhancements in the vicinity of these objects. At present plasmonics
has been the subject of extensive theoretical and experimental
research and experiences a dramatic growth of applications  \cite{maier,pelton,nie,prodan}.
Recently an interesting experiment has been reported based on a
connection of both fields: high-order harmonic generation (HHG) has
been demonstrated by nJ pulses directly from a laser oscillator by
exploiting the local field enhancement near a metallic bowtie-shaped
gold nanoelement \cite{kim}. The elimination of complex and expensive
amplifiers and the opportunity to perform HHG and attosecond
experiments at MHz repetition rates can significantly extend the
ability of extreme VUV generation and attosecond pump-probe
spectroscopy and will lead to novel applications in lithography and
imaging. However, a theoretical
description of plasmon-enhanced HHG, which is a key 
 for further progress in this field, is currently missing.

In this paper we present a theoretical investigation of
plasmon-enhanced HHG in a noble gas in the vicinity of a silver
nanocone and a silver bowtie structure. Our goal is directed to an
understanding of the role of metal nanostructures and the
inhomogeneous plasmonic field enhancement in the HHG process. We
describe the microscopic laser-atom interaction by the
time-dependent dipole moment in the semi-classical approximation \cite{lewenstein}.
In the vicinity of the metal nanostructures the input laser intensity 
is enhanced by up to three orders of magnitude by the plasmons in
so-called "hot spots" and sensitively depends on the position. This
strongly enhanced field ionizes the noble gas atoms, the resulting
electrons are accelerated in the enhanced field and
recollide with the parent ion,
 transforming their kinetic energy into
 high harmonic radiation. In the hot spot, electron experience the
inhomogeneous electric field, and some of
them will hit the metal surface and never return to the parent ion.
We show that both field inhomogeneity and the electron absorption at
the metal surface lead to the emission of even harmonics as well as
to more than a twofold increase of the
 cut-off frequency.

We describe HHG in the framework of a model that is an extension of
the Lewenstein model modified to account for the field inhomogeneity
in the vicinity of the nanostructure, as well as for the electron
absorption by the metallic surface. Under the action of the electric
field $E(t)$,
 the electron leaves the atom at the time moment $t_s$ and later
recombines with the parent ion at the time $t_f$. 
The maximum HHG contribution is
provided by electrons with the canonical momentum $p$
for which their position at $t_f$ coincides with that of the parent ion.
Neglecting the Coulomb
potential and using the stationary phase method\cite{lewenstein}, the time-dependent high-harmonic
dipole moment is given by 
\begin{eqnarray}
&&x(t_f)=i\frac{e}{2\omega_0^{5/2}m_e}\int_{-\infty }^{t_f}\left( \frac{\pi }{\epsilon +i\Delta t/2}%
\right)^{3/2}H(t_f,t_s)\nonumber \\ 
&&\times d_{x}(p(t_{s})-eA(t_s))d_{x}^{\ast}(p(t_f)-eA(t_f))\nonumber \\ 
&&\times E(t_f)\exp (-i\frac{S(t_f,t_{s})}\hbar)dt_s+c.c.  \label{1}
\end{eqnarray}%
Here $\epsilon$ is an arbitrary small parameter, $\omega_0$ is the central frequency,
$d_{x}(p)=i2^{7.25}[\hbar\omega_0m_e^2I_p]^{5/4}\pi^{-1}p/(p^2+\alpha)^3$ with
$\alpha=2m_e^{1/2}I_p^{1/2}$
is the dimensionless dipole matrix element, and $S(t_f,t_s)=
\int_{t_{s}}^{t_f}(I_{p}+(p(t)-eA(t))^{2}/2m_e)dt$ is the classical action equal
in the Lewenstein model to $S_0(t_f,t_s)=I_{p}\Delta
t-0.5e^2(\Delta B^2/\Delta t+\Delta C)/m_e$.
 Here $\dot{A}\equiv dA(t)/dt=E(t)$, $\dot{B}(t)=A(t)$, $\dot{C}(t)=A^2(t)$, for any function $F$ we 
define $\Delta F\equiv
F(t_f)-F(t_s)$, $\Delta t= t_f-t_s$, and $I_{p}$ is the
ionization potential. 
 For the unmodified Lewenstein
model, in the stationary-phase approximation the canonical momentum is given by $e\Delta
B/\Delta t$.
$H(t_f,t_s)$ is a function describing the electron
absorption at the metal surface; in the unmodified Lewenstein model we have $%
H(t_f,t_s)\equiv 1$. We  extend the semiclassical HHG model
 accounting for the inhomogeneity of the
field in the hot spot by adding the
first-derivative term to the field $E(t,x)=E(t)(1+x/d_{inh})$, and
consider this derivative term as a perturbation, including only the
first-order correction to the electron
trajectory. As a result, the expressions for the momentum $p(t_s)$ and $%
S(t_f,t_s)$ are modified as follows:
\begin{eqnarray}
&&\!\!\!\!\!\!p(t_s)=e[A(t_s)+  \nonumber \\
&&\!\!\!\!\!\!\!\frac{\Delta B-A(t_{s})\Delta t+\beta (0.5(\Delta B)^{2}-\Delta
D+C(t_{s})\Delta t)}{\Delta t-\beta(2\Delta G+\Delta t[B(t)+B(t_{s})])}],
\end{eqnarray}
\begin{eqnarray}
&S(t_f,t_{s})&=S_{0}(t_f,t_s)+e^2\beta m_e^{-1}
\{p(t_s)^{2}[2\Delta G \nonumber \\ &&+(B(t_f)+B(t_{s}))\Delta t  
]+p(t_{s})[-\Delta C\Delta t  \nonumber \\
&&+2\Delta D]+(C(t_f)+C(t_{s}))\Delta B-2\Delta F\},
\end{eqnarray}%
where $\dot{D}(t)=C(t)$, $\dot{F}(t)=C(t)E(t)$, $\dot{G}(t)=B(t)$,
 $\beta =e/(m_{e}d_{inh})$, and $T_{0}$ is the optical period.
Another modification of the model is described by the function $H(t,t_{s})$
under the integral in Eq.~(1). This function is equal to $1$ unless the
electron during the motion hits the metal surface positioned at $d_{sur}$,
otherwise we assume that the electron is absorbed by the surface and set $%
H(t,t_{s})=0$. We have used the finite-element Maxwell solver
JCMwave to model the plasmonic field enhancement in the vicinity of
the nanostructure, taking the experimental complex-valued dielectric
function of silver. For 
short input pulses, to include the dispersion of silver and
of the field distribution near the
nanostructure, we calculated the field
distribution for a set of wavelengths within the input pulse
spectrum and then reconstructed the position-dependent pulse profile
from the wavelength- and position-dependent field. In this
paper, we disregard the modification of the
field induced by the neighboring structure, the
depolarization effects are weak in our case.
To prove that the mesh size is small enough, we halved it  
and found no significant change in the results.

\begin{figure}[b]
\includegraphics[width=0.45\textwidth]{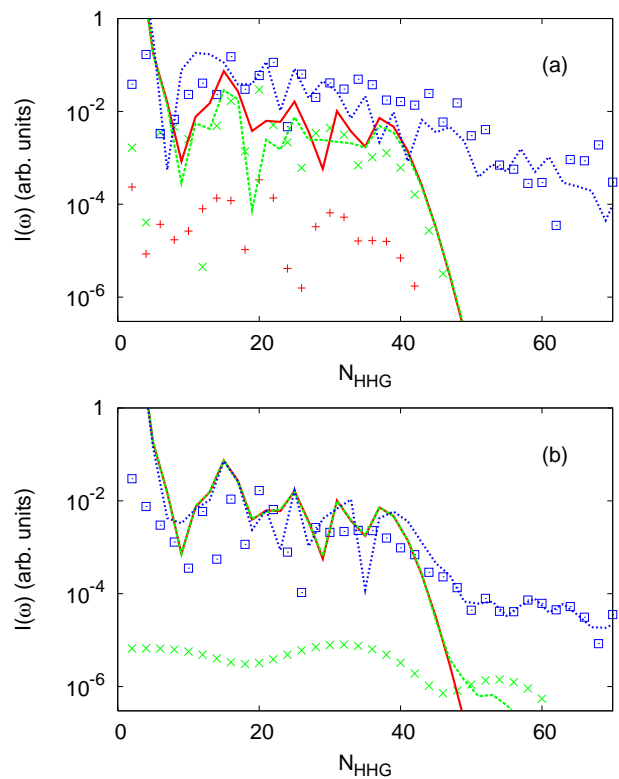}
\caption{Influence of the field inhomogeneity (a) and of the metal
surface (b) on the HHG process. Odd harmonics are presented by
curves while even harmonics are shown by points. In (a), the field
has the inhomogeneity scale $d_{inh}$ of 1000 nm (red solid curve
and red vertical crosses), 20 nm (green short-dashed curve and green
diagonal crosses), and 5 nm (blue long-dashed curve and blue
squares). In (b), the distance to the metallic surface $d_{sur}$ is
3 nm (red solid curve), 2.62 nm (green short-dashed curve and green
diagonal crosses), and 1.5 nm (blue long-dashed curve and blue
squares). For both cases, cw radiation at 830 nm with 
 intensity in the hot spot of 200 TW/cm$^2$ in argon is considered.}
\label{th}
\end{figure}

First we analyze the principal effects described by the above extended
model in dependence on the model parameters $d_{inh}$
 and $d_{sur}$. To understand the effect of the
inhomogeneity of the field and of the metal surface on the HHG process, in
Fig. 1 we consider the HHG spectra for different 
 parameters $d_{inh}$ [Fig. 1(a)] and 
 $d_{sur}$ [Fig. 1(b)] for cw radiation with
a typical intensity in the hot spot of 200 TW/cm$^{2}$. In Fig. 1(a) even
for a very low inhomogeneity, characterized by $d_{inh}=1000$ nm,
the inverse spatial symmetry is broken, which leads to a
qualitatively new effect: the generation of even harmonics (shown by
the blue crosses). For larger inhomogeneities with $d_{inh}= 20$ nm
the amplitude of the even harmonics approaches that of odd
harmonics. Simultaneously, the harmonic cutoff shifts to higher
harmonic numbers and become much less pronounced. Due to
inhomogeneity, electrons have higher velocities when they reach the
parent ion, leading to a higher kinetic energy 
and therefore to a higher cut-off frequency. The
influence of the metal surface and the associated electron
absorption is studied in Fig. 1(b). One can see that for a distance
to the surface $d_{sur}=3$ nm, no electron reaches the surface, and
correspondingly no modification in the dipole moment and no even
harmonics appear. For shorter $d_{sur}$, some electrons will reach
the
 surface. Since the function $H(t,t_{s})$
introduces a sharp modulation in the temporal profile of the
harmonics, an extension of the spectrum to higher harmonic numbers
can be seen in Fig. 1(b), combined with the emission of even
harmonics.
From Fig. 1 one can see that the combined effect of the field
inhomogeneity and of the metal surface can increase the
 HHG cutoff by a factor larger than two.

\begin{figure}[t]
\includegraphics[width=0.45\textwidth]{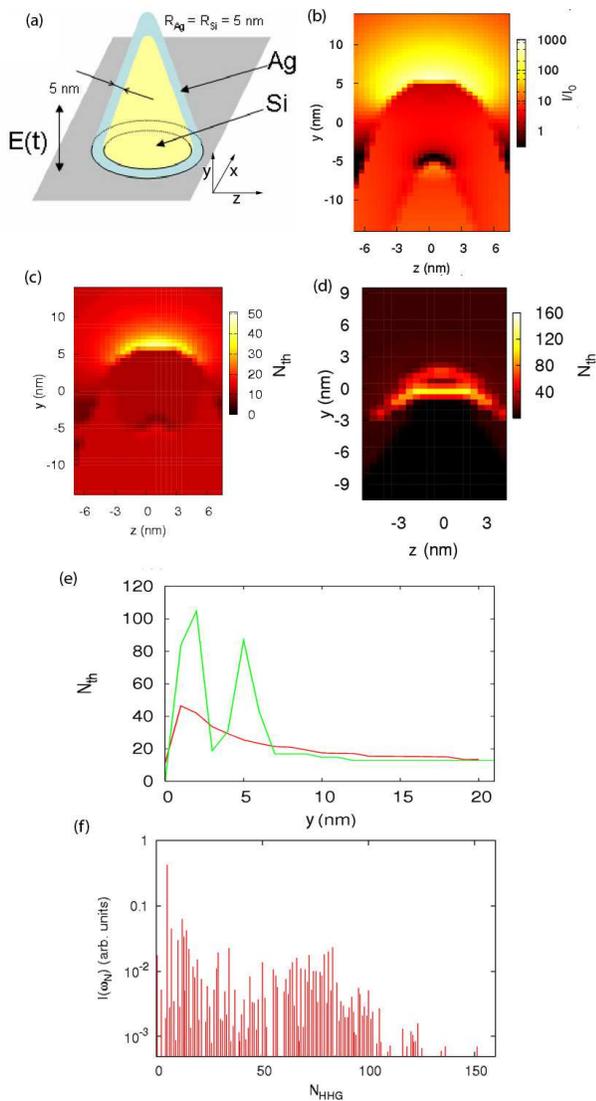}
\caption{High harmonic generation in the vicinity of a nanocone
structure. In (a), the scheme of the nanocone is shown.
 The curvature radii and the thickness of silver
coating are 5 nm, the $y$-polarized field
propagates along the $x$ direction. In (b) the intensity enhancement is
presented for the $zy$ cross-section of the incident field at 830 nm. In
(c) and (d), the distribution of the harmonic cutoff is shown for
the model without (c) and with (d) modifications by the field
inhomogeneity and metal surface for 10-fs input pulses with 0.3
TW/cm$^2$ intensity and argon surrounding the tip. In (e) the cutoff
for the modified (green curve) and unmodified (red curve) model is
shown along the $y$ axis, and in (e) the spectrum for cw excitation
with the above parameters is shown.} \label{nc}
\end{figure}

Next we perform a realistic simulation of HHG in the vicinity of a
silver nanocone as depicted in Fig. 2(a).  Such nanocone structure
can be produced by silver deposition on a chemically created silicon
needle without the need of nanoscale lithographic control. The
curvature radius of the nanostructure edges is the key
parameter determining the maximum field enhancement, with optimum structure having the smallest curvature radius.
 We have assumed
radii of 5 nm, in correspondence with the current state of the
manufacturing technique. The spatial field distribution shown in
Fig. 2(b) with parameters given in the caption demonstrates intensity
enhancement factors of up to 1000 in the "hot spot" in the vicinity
of the tip. From the comparison of $xy$ and $zy$ cross-sections (not
shown) one can see that the field enhancement is higher on the
surface directed towards the incoming beam, but this is a minor
effect compared to the strong enhancement near the tip. Note that
the field is inhomogeneous on the scale below 10 nm in the "hot
spot" which is located directly above the metal surface. The
parameters $d_{inh}(x,y,z)$ and $d_{sur}(x,y,z)$ are calculated from
this spatial distribution of the field and are used for the HHG
simulation by using the extended model as given by Eq. (1)-(3). In
Fig. 2(c) and (d) we present the spatial distribution of the
harmonic cutoff $N_{th}$ for an excitation by 830-nm sech-shaped pump
pulses with intensity of only 0.3 TW/cm$^{2}$, FWHM of 10 fs, and
polarization and wavevector as indicated in Fig. 2(a). The
field enhancement in the "hot spot" increases the intensity to
roughly 300 TW/cm$^{2}$, which for the chosen wavelength corresponds
in the unmodified Lewenstein-model to a
 cut-off of roughly the 50$^{th}$ harmonic, as can be seen in
Fig. 2(c). However, due to the effect of the field inhomogeneity and the
presence of the metal surface, the cutoff increases up to 105, which
correspond to the emission of 150-eV photons using an input
intensity of only 0.3 TW/cm$^{2}$. Such pulses can be obtained
from a typical laser oscillator without amplifier. In Fig. 2(e)
 a comparison between the extended model including
the effects of the field inhomogeneity and metal surface, and the unmodified
Lewenstein model is drawn. The cut-off harmonic number as
a function of the position reaches values of above 100 for the modified
model, while the unmodified only predicts values around 40. Further from the
tip of the nanocone, the field become relatively homogeneous, and
the distance to the metal surface is large, therefore both models
coincide. Note however, that although the modifications of the
Lewenstein model are important only in a small fraction of the volume, it
is exactly this volume ("hot spot") which dominate the generation of high
harmonics and therefore these modifications become
especially important. This is clearly visible in Fig. 2(f), where
we show the spectrum of the high harmonics emitted in the $x$ direction
integrated over the nanostructure position,
where a significant emission of  even harmonics as well as the
extended and smoothed cutoff can be seen.
We have also calculated the phases of the HHG harmonics separately for short and long elelctron trajectories,
and found that the modification of phases due to the field inhomogeneity is negligible for
short trajectories but reaches $\pi/2$ for the long ones. This, however, will not influence the divergence of
the HHG radiation, since the hot spot is smaller than the typically
considered wavelength and is a pointlike source. 
In \cite{supp} we calculated the variation of the high-harmonic phases 
 with different intensities depending on the transverse
positions in the beam. It is shown that in the considered example
the phases change only slightly for different intensities within
the beam for harmonic numbers below 50, which can be compensated by an
appropriate curvature of the pump beam. For higher
harmonic numbers the phases show strong changes, resulting in beam distortion
due to irregular spatial modulation.

\begin{figure}[t]
\includegraphics[width=0.375\textwidth]{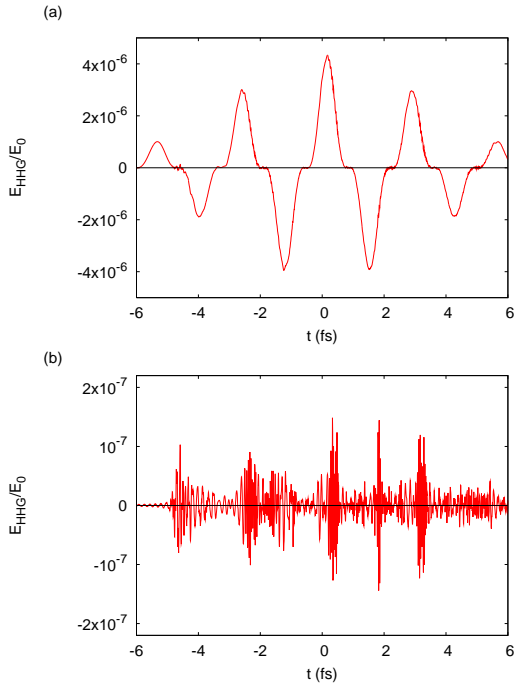}
\caption{Temporal shape (a) and the temporal shape after filtering
out low harmonics ($N<$40) of the output(b) with input intensity of
0.3 TW/cm$^2$ and duration of 10-fs at 830 nm for a nanocone
structure.}
\end{figure}

In Fig. 3, the temporal profile of the output harmonic radiation is
shown for 10-fs input pulses. The temporal profile in Fig. 3(a) is
dominated by low-order ($<$10) harmonics. It is well-known that
filtering out the lower harmonics leaves only few electron
trajectories with maximum kinetic energy at the recollision time,
which can lead to the formation of attosecond pulses. In Fig. 3(b)
we present the temporal profile for the above case but with
harmonics below the 50th order filtered out. One can see the
formation of few pulses with down to 150-attosecond FWHM.
These pulses are shorter and more irregular than those
obtained from the unmodified model (not shown).

\begin{figure}[t]
\includegraphics[width=0.45\textwidth]{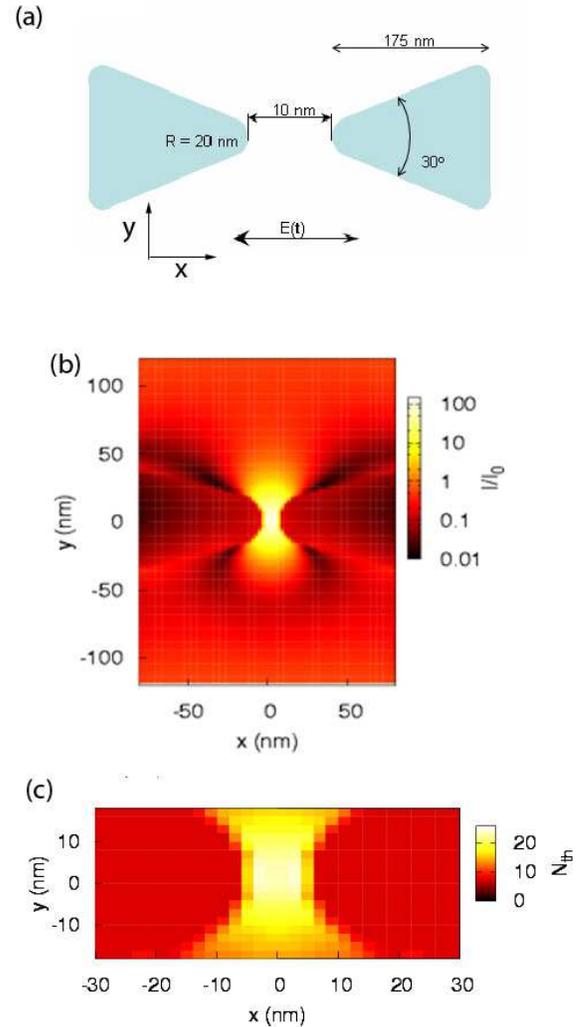}
\caption{High harmonic generation in the vicinity of a bowtie
structure. In (a), the scheme of the bowtie structure and its
geometric parameters are shown, the $x$-polarized field propagates
along $y$ direction. In (b) the field enhancement is presented for
the $xy$ cross-section of the incident field at 830 nm; in (c), the
distribution of the harmonic cutoff is shown for 10-fs input pulses
with 0.5 TW/cm$^2$ intensity for HHG in argon surrounding the bowtie
structure.} \label{bt}
\end{figure}

In order to compare the results of our model with experimental
measurements, we have simulated high harmonic generation in the vicinity of
a bowtie nanoantenna, such as used in Ref. \cite{kim}. Structure and
parameters of the studied bowtie antennas is shown in Fig. 4(a)
consisting of two silver isosceles triangles with curvature
radius and distance between the vortices in the range of 10 nm.
In Fig. 4(b), the field enhancement by the bowtie antenna
 with parameters as given in the figure is presented, which shows a
"hot spot" between the inner vortices of the triangles and a
maximum enhancement of about 10$^{2}$. For other geometrical parameters of
the structure, the enhancements drops with increasing curvature
radius and distance between the triangle tips. By using this field
enhancement HHG is calculated by using the extended model. In Fig.
4(c) for incident 10-fs pulses with intensity of 0.5 TW/cm$^{2}$ the spatial
distribution of the threshold harmonic number $N_{th}$ is presented. The
calculated maximum harmonic number is 23, which corresponds to the
wavelength of 36 nm. This is in good agreement with the
experimental findings of \cite{kim}, where harmonics down to 45 nm were
observed.
Note that both the geometry of
the bowtie antenna and the input intensity are not exactly defined in \cite%
{kim} which can lead to deviations between theory and the
experiment.
In contrast to the nanocone, in the case of the bowtie antenna a relatively low
field enhancement due to larger curvature radii leads to a smaller amplitude of
electron motion which, together with a relatively
homogeneous field distribution, makes the model modifications less
important than in the case of the considered nanocone. Our calculations 
 predict that the cutoff value has increased by only roughly
20\% due to the effects of the inhomogeneity and the metallic surface.


\end{document}